\documentclass[aps, prb, reprint, floatfix, citeautoscript]{revtex4-1}
    \usepackage{amsmath}
    \usepackage{amssymb}
    \usepackage{graphics}
    \usepackage{graphicx}
    \usepackage{color}
    \usepackage{physics}    
    \usepackage[colorlinks=true, citecolor=blue, urlcolor=blue, linkcolor=black]{hyperref}
    \usepackage[capitalize]{cleveref}
    \usepackage[detect-weight=true, detect-family=true, separate-uncertainty=true, multi-part-units=single]{siunitx}

\begin{document}
        \title{On the origin of the controversial electrostatic field effect in superconductors}
        \author{I.~Golokolenov}
        \altaffiliation[Also at: ]{P.\,L.\,Kapitza Institute for Physical Problems of RAS, 2 Kosygina St., Moscow, 117334, Russia;
        \newline National Research University Higher School of Economics, Moscow 101000, Russia}
        \author{A.~Guthrie}
        \author{S.~Kafanov}\email{sergey.kafanov@gmail.com}
        \author{Yu.\,A.~Pashkin}\email{y.pashkin@lancaster.ac.uk}
        \author{V.~Tsepelin}
        \affiliation{Department of Physics, Lancaster University, Lancaster, LA1 4YB, United Kingdom}

        \maketitle
        \textbf{In semiconductor electronics, the field-effect refers to the control of electrical conductivity in nanoscale devices \cite{Datta_ElectronicTransport}, which underpins the field-effect transistor, one of the cornerstones of present-day semiconductor technology \cite{SzeNg_PhysicsSemiconductorDevices}. The effect is enabled by the penetration of the electric field far into a weakly doped semiconductor, whose charge density is not sufficient to screen the field. On the contrary, the charge density in metals and superconductors is so large that the field decays exponentially from the surface and can penetrate only a short distance into the material. Hence, the field-effect should not exist in such materials \cite{Ashcroft_SolidStatePhysics}. Nonetheless, recent publications have reported observation of the field-effect in superconductors and proximised normal metal nanodevices \cite{NatNano.13.802(2018), NanoLett.18.4195(2018), ACSNano.13.7871(2019), PhysRevApplied.11.024061(2019)}. The effect was discovered in gated nanoscale superconducting constrictions as a suppression of the critical current under the application of intense electric field and interpreted in terms of an electric-field induced perturbation propagating inside the superconducting film. Here we show that ours, and previously reported observations, governed by the overheating of the constriction, without recourse to novel physics. The origin of the overheating is a leakage current between the gate and the constriction, which perfectly follows the Fowler-Nordheim model of electron field emission from a metal electrode \cite{ProcRoyalSocA.119.173(1928)}.
        }
        
        \begin{figure*}
            \includegraphics[width=\linewidth]{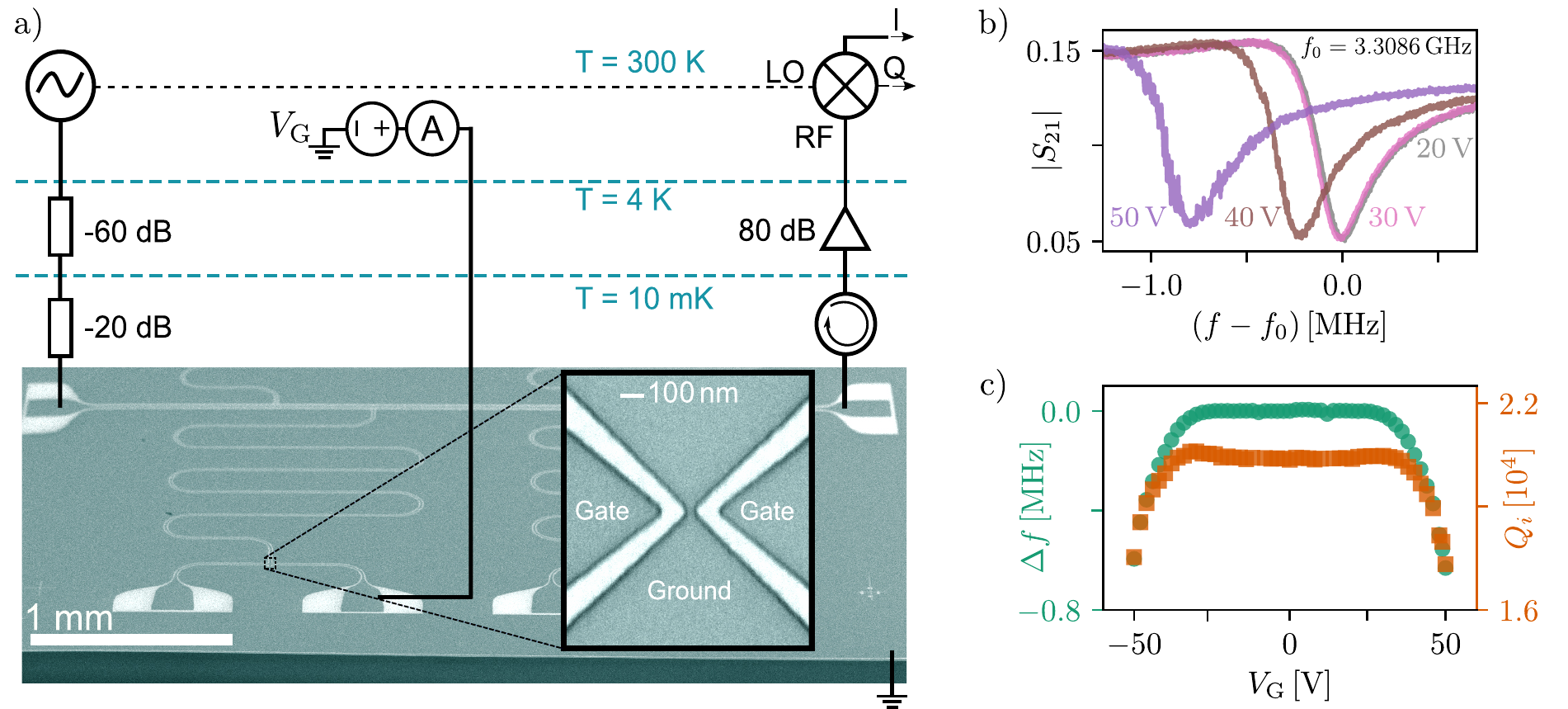}
            \caption{(Colour online) \textbf{Experimental details and basic sample characterisation.} \textbf{(a)} Schematic of the measurement setup with an electron micrograph of the investigated superconducting structure consisting of a coplanar transmission line with four capacitively coupled quarter wavelength microwave resonators. Each resonator is terminated to the ground plane by the constriction (Dayem bridge) shown in the inset.
            \textbf{(b)} A collection of the resonance curves at different gate voltages, \(V_\mathrm{G}\), for an investigated resonator. Note that the resonance frequency rapidly decreases at large gate voltages, \(|V_\mathrm{G}|>\SI{30}{\volt}\). \textbf{(c)} The shift of the resonance frequency (left axis) and the internal quality factor of the resonator (right axis) as a function of applied gate voltage.}
            \label{fig:Scheme&Resonances}
        \end{figure*}
        
        Controlling the properties of superconducting films using gates was first proposed in the 1960\,s when it was found that electrostatic charging can affect the superconducting transition temperature of thin tin and indium films \cite{PhysRevLett.5.248(1960)}. Several years later, a mesoscopic superconductor–normal metal–superconductor Josephson junction was realised by controlling the supercurrent flow via a ‘‘normal’’ current traversing the normal metal between the superconducting electrodes \cite{ApplPhysLett.72.966(1998)}. This control was attributed to the modified quasiparticle distribution which was driven far from equilibrium by a voltage applied across the normal metal \cite{PhysRevLett.81.1682(1998)}.
        
        Another technique for controlling the Josephson supercurrent is to introduce a semiconductor in which carrier concentration can be tuned by electrostatic effects. As a result, a Josephson Field-Effect Transistor (JoFET) was realised by building small hybrid superconductor--semiconductor structures where a region of sub-micron-long high-mobility two-dimensional electron gas (2DEG) was in good galvanic contact with two superconducting electrodes \cite{JApplPhys.51.2736(1980)}. Through the development of nanofabrication techniques, it became possible to build such structures using various semiconductors and superconductors. A supercurrent through the whole structure was observed and controlled electrostatically by a nearby gate, due to the proximised superconductivity in the semiconductor \cite{PhysRevLett.54.2449(1985), IEEEElectDevLett.6.297(1985), ApplPhysLett.55.1909(1989)}. While at low voltages these devices act as Josephson junctions with a gate-controlled critical current, at high voltages they behave as conventional FETs. 
        
        In the later experiments, the 2DEG was replaced by indium arsenide semiconductor nanowires, with aluminium-based superconducting electrodes \cite{Science.309.272(2005)}. Below \SI{1}{\kelvin}, due to the proximity effect, the nanowires form superconducting weak links operating as mesoscopic Josephson junctions with electrically tunable coupling. A gate voltage controls the electron density in the nanowire, and regulates the supercurrent. Finally, the availability of semiconductor graphene flakes resulted in hybrid graphene/superconductor devices  where the gate voltage controls supercurrent via either quasiparticles in the conduction band or by quasiholes in the valence band \cite{Nature.446.56(2007)}. 
        
        While the field-effect in hybrid semiconductor-superconductor structures was predicted \cite{JApplPhys.51.2736(1980)} and confirmed experimentally \cite{PhysRevLett.54.2449(1985), IEEEElectDevLett.6.297(1985), ApplPhysLett.55.1909(1989)}, it is not expected to exist in all-superconducting devices where the high carrier density screens the applied electric field. Therefore the observation of the electrostatic field-effect in metallic nanostructures \cite{NatNano.13.802(2018), NanoLett.18.4195(2018), ACSNano.13.7871(2019), PhysRevApplied.11.024061(2019)} warrant further studies and interpretation.
         
        Here we investigated gated superconducting constrictions (see inset in \cref{fig:Scheme&Resonances}(a)), identical to the structures described in references \cite{NatNano.13.802(2018), NanoLett.18.4195(2018), ACSNano.13.7871(2019), PhysRevApplied.11.024061(2019)}, in order to understand the origin of the observed field-effect. The structure in the inset of \cref{fig:Scheme&Resonances}(a) is a Dayem bridge accompanied by two control gates located \SI{100}{\nano\meter} apart on both sides of the constriction. The devices were formed in a \SI{30}{\nano\meter} thick superconducting vanadium film (\(T_c = \SI{4.18}{\kelvin}\))  on the surface of oxidised undoped silicon. Since the constriction has characteristic dimension comparable with the vanadium superconducting coherence length \(\xi_0\approx\SI{40}{\nano\meter}\), \cite{J.PhysicsConf.592.012137} it is a Josephson junction with inductance \(L_\mathrm{J}=\hbar/(2e I_c)\), where \(I_c\) is the superconducting critical current through the constriction.

        \Cref{fig:Scheme&Resonances}(a) shows the schematic of our experiment and electron micrograph of the sample. To investigate the effect of the gate voltage on the superconductors, we have embedded each Dayem bridge into the current antinode of the quarter-wavelength coplanar microwave resonator, formed in the same vanadium film. A series of four meander-shaped resonators are incorporated into a manifold frequency multiplexing network (FMN) \cite{Mohsin_ManifoldMultiplexer}, which allows independent probing of each resonator at its resonance frequency using a single feedline. The coupling strength of each resonator to the feedline is weak, such that the quality factor is determined only by the resonators' internal losses.
         
        The experiments were performed in a cryogen-free dilution refrigerator with a base temperature of \SI{10}{\milli\kelvin}. The incoming microwave tone was filtered and attenuated at each temperature stage of the cryostat. After passing through the FMN, the transmitted signal was amplified by a series of cryogenic amplifiers and detected using an \(\mathrm{IQ}\) demodulator, which allowed independent measurements of the in-phase, \(\mathrm{I}\), and quadrature, \(\mathrm{Q}\). The voltage on the control gates was applied through DC lines filtered at \SI{10}{\milli\kelvin}. We used a biasing scheme to measure the leakage current, \(I_\mathrm{L}\), and differential conductance of the gap between the gate and the constriction, \(\dd I_\mathrm{L}/\dd V_\mathrm{G}\).
        
        All four resonators embedded into the FMN behaved alike, so here we present experimental results for one device with a resonance frequency \(f_0=\SI{3.3086}{\giga\hertz}\) measured at zero gate voltage. The resonance frequency, \(\omega=(LC)^{-1/2}\), of the resonator is determined by its capacitance, \(C\), and inductance, \(L\). The internal \(Q\)-factor is given by \(Q_i=\omega L/R\), where \(R\) represents dissipation in the coplanar waveguide forming the resonator \cite{DiPaolo_NetworksDevices}. The total inductance is the sum of the resonator's geometric inductance, \(L_g\), and the Josephson inductance of the Dayem bridge. In our resonators the geometric inductance is much larger than the Josephson inductance, \textit{i.e.}, \(L_g\gg L_\mathrm{J}\). 
        
        \Cref{fig:Scheme&Resonances}(b) shows a collection of the resonance curves measured at various gate voltages, \(|V_\mathrm{G}|<\SI{50}{\volt}\), \textit{i.e.}, the same voltage range as has been used in the reported experiments \cite{NatNano.13.802(2018), NanoLett.18.4195(2018), ACSNano.13.7871(2019), PhysRevApplied.11.024061(2019)}. The resonance frequency exhibits little change up to \(|V_\mathrm{G}|\approx\SI{25}{\volt}\), however, at higher \(V_\mathrm{G}\) it significantly decreases. The effect can be explained by the critical current suppression in the Dayem bridge, which leads to an increment of \(L_\mathrm{J}\) and reduction of the resonance frequency, since:
        \begin{equation}
            \omega\approx\dfrac{1}{\sqrt{L_g C}} \left(1-\dfrac{1}{2}\dfrac{L_\mathrm{J}}{L_g}\right).
        \end{equation}
        The shift of the resonance frequency is bipolar and symmetric around zero gate voltage, see \cref{fig:Scheme&Resonances}(c, left axis). Our observations confirm the suppression of supercurrent in the Dayem bridge by the gate voltage \cite{NatNano.13.802(2018), NanoLett.18.4195(2018), ACSNano.13.7871(2019), PhysRevApplied.11.024061(2019)} but disagree with the existence of the electrostatic field-effect for the reasons presented below.
        
        \begin{figure*}
    	    \includegraphics[width=\linewidth]{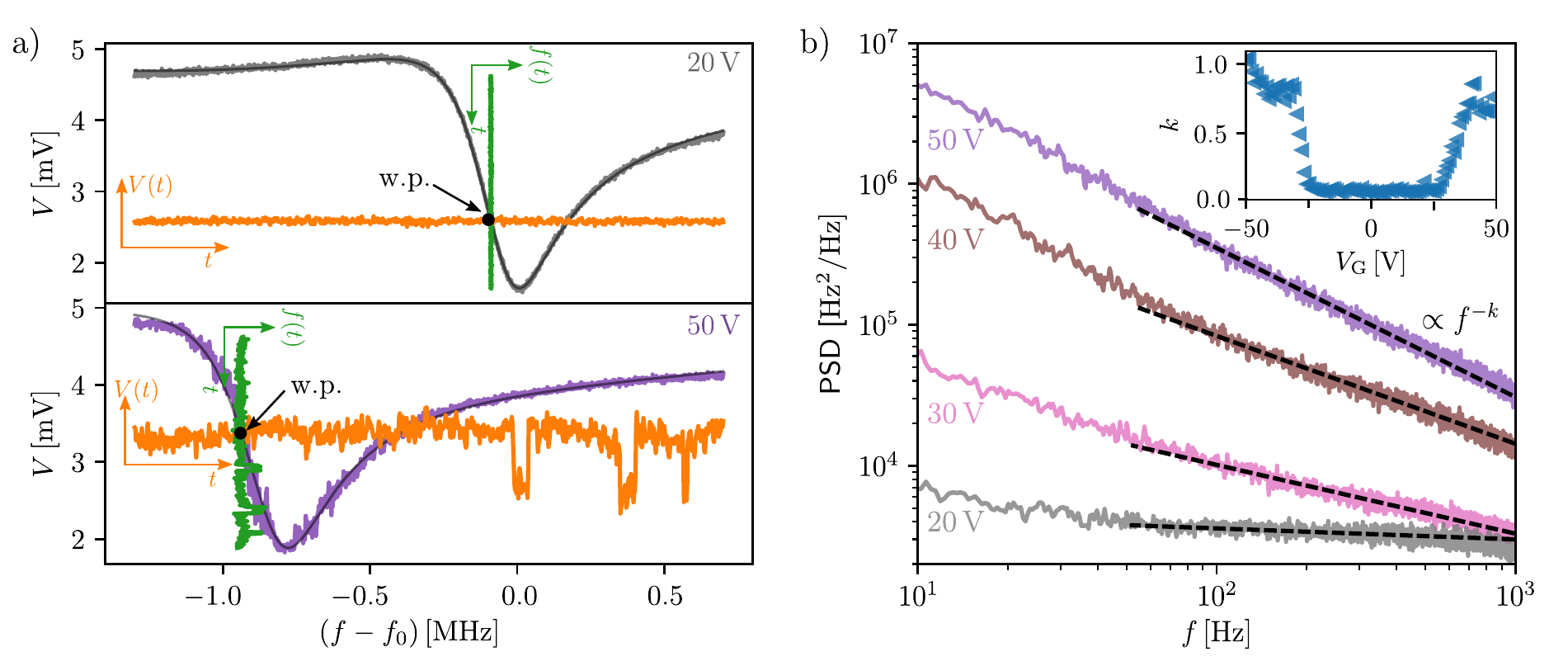}
    	    \caption{(Colour online) \textbf{Noise properties of the superconducting coplanar resonator.} \textbf{(a)} Principle of the frequency noise measurements. Examples of resonance curves at \SI{20}{\volt} (top panel) and \SI{50}{\volt} (bottom panel) gate voltage applied to the Dayem bridge. At each gate voltage, the steepest point of the magnitude of the resonance is used as the working point (w.\,p.) for time trace measurements. Examples of voltage and frequency fluctuations with duration of \SI{1}{\second} are shown in orange and green colours, respectively. \textbf{(b)} Power spectral density (PSD) of fluctuations at various gate voltages. Dashed lines are the fits of PSD functions by \(\propto f^{-k}\). Inset shows the dependence of the fitted exponent \(k\) on the gate voltage at \(\SI{10}{\milli\kelvin}\).
    	    \label{fig:Noise}}
        \end{figure*}   
        
        \Cref{fig:Scheme&Resonances}(c, right axis) demonstrates that the application of gate voltages within \(|V_\mathrm{G}|\approx\) \SI{25}{\volt} produces an  expected  small rise of the \(Q\)-factor caused by an increment of the Josephson inductance 
        \begin{equation}\label{Qfactor}
            Q_i\approx\dfrac{1}{R}\sqrt{\dfrac{L_g}{C}}
            \left(1+\dfrac{1}{2}\dfrac{L_\mathrm{J}}{L_g}\right).
        \end{equation}
        However, \(Q_i\) drops rapidly at higher voltages, which can only be caused by the increase of the internal losses of the resonator, \textit{i.e.}, dissipation, \(R\). Note, that the bipolar and symmetric gate dependency of the \(Q\)-factor is unexpected by itself for the for negatively charged superconducting carriers, \textit{i.e.}, Cooper pairs. Furthermore, the resonance curves become visibly noisier at high gate voltages, which also cannot be attributed to the stationary change of a reactive parameter, such as the Josephson inductance. Hence, all observations point towards higher dissipation in the resonator at large applied gate voltages and cast doubt on the detection of a straightforward field-effect in superconducting constrictions.

        \begin{figure*}
            \includegraphics[width=\linewidth]{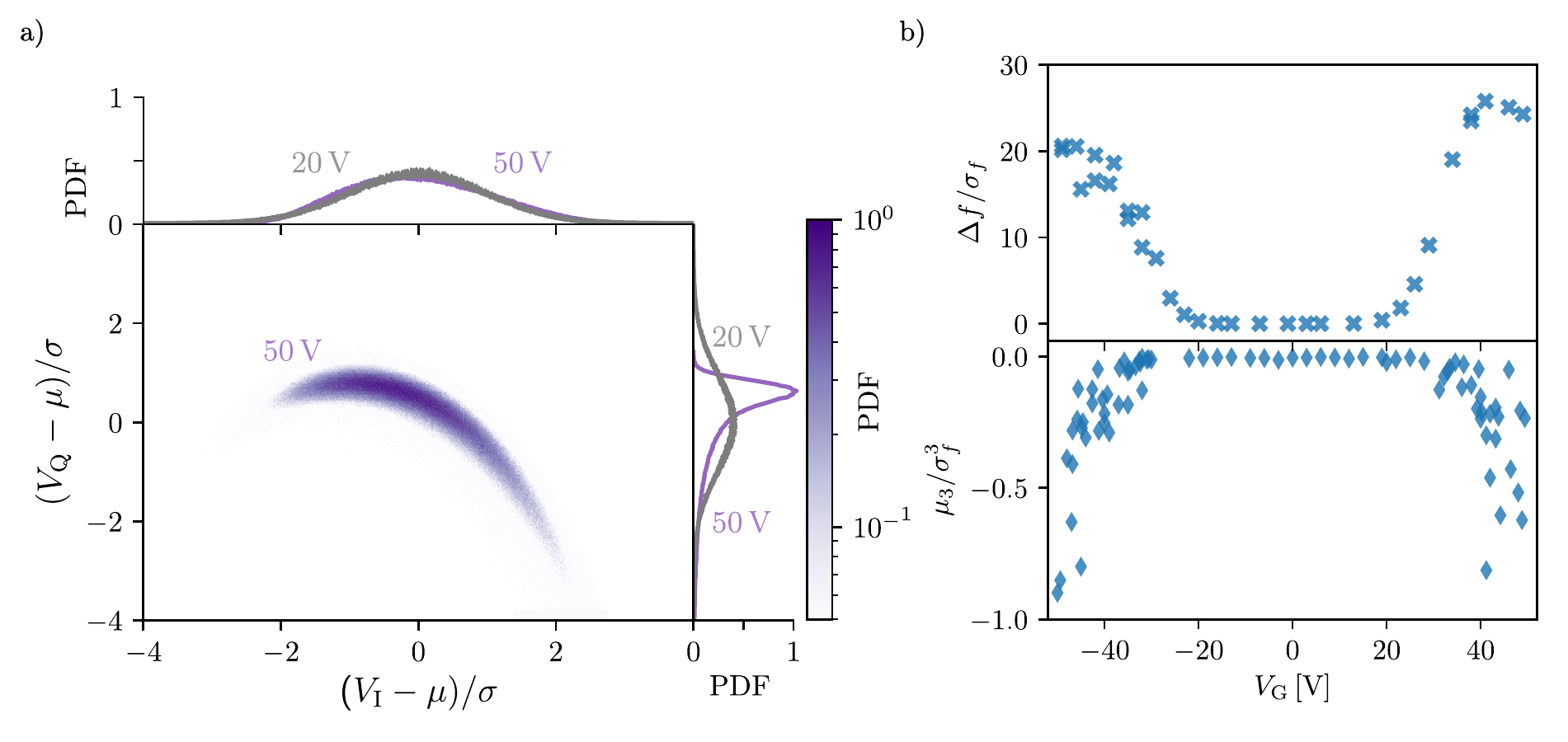}
            \caption{(Colour online) \textbf{Histogram analysis of the voltage noise and reconstructed noise of the resonance frequency at \SI{10}{\milli\kelvin}.} \textbf{(a)} An example of two-dimensional probability density function (PDF) of a normal deviate of the voltage noise obtained from the homodyne detection of the transmitted signal at frequency corresponding to w.p. in \cref{fig:Noise}(a) for \(V_\mathrm{G}=\SI{50}{\volt}\). Side insets show the corresponding one-dimensional PDFs for in-phase and quadrature components of the voltage noise, in contrast to the cognate PDFs obtained at \(V_\mathrm{G}=\SI{20}{\volt}\). \textbf{(b)} The ratio of the change of the resonance frequency to the variance of the frequency fluctuations. The dependencies of the normalised frequency shift and standardised 3\textsuperscript{rd}-moment (skewness) of the frequency fluctuations PDF from \(V_\mathrm{G}\).}
            \label{fig:histograms}
        \end{figure*}
        
        We investigated the noise properties of the resonator in the same range of the applied gate voltages to understand the nature of the dissipation. The principle of the noise measurements is presented in \cref{fig:Noise}(a). The top and bottom panels show examples of the resonance curves for the magnitude of the transmitted microwave signal, for two gate voltages of \SI{20}{\volt} and \SI{50}{\volt}, correspondingly. After selecting the frequency of the working point (w.\,p.), \textit{viz.}, at the steepest point of the resonance curve, we drove the resonator at this frequency and detected the transmitted signal as a function of time. \Cref{fig:Noise}(a) depicts an example of the measured time traces of the magnitude of the transmitted signal \(V(t)\) in orange with a clear growth of fluctuations with larger \(V_\mathrm{G}\). The signatures of the telegraph-like noise visible at \SI{50}{\volt} time trace start to appear at \(|V_\mathrm{G}|\approx\SI{35}{\volt}\). The appearance and growth of the low-frequency telegraph noise with \(V_\mathrm{G}\) contradict the claim \cite{NatNano.13.802(2018), NanoLett.18.4195(2018), ACSNano.13.7871(2019), PhysRevApplied.11.024061(2019)} of static suppression of superconductivity by the electric field.
        
        To compare our measurements with other possible measurements of the supercurrent noise, we converted the magnitude time traces to the traces of the frequency fluctuations \(f(t)\) (green traces in \cref{fig:Noise}(a)), by dividing the former by a gradient of the resonance curve at the working point. \Cref{fig:Noise}(b) shows examples of the frequency noise spectra for four distinct gate voltages. At low \(V_\mathrm{G}\), the noise spectrum is almost frequency independent, \textit{i.e.}, white noise. However, as \(V_\mathrm{G}\) increases the noise changes both quantitatively and qualitatively, and, importantly, it grows faster at low frequencies such that the slope of the frequency dependence increases with \(V_\mathrm{G}\). The noise of the resonance frequency is associated with the noise of the Josephson inductance of the Dayem bridge, which is proportional to the critical current noise \cite{PhysRevLett.76.3814(1996), PhysRevB.53.R8891(1996)}. Under the assumption that the application of \(V_\mathrm{G}\) does not change the temperature of the nanoconstriction \cite{NatNano.13.802(2018), NanoLett.18.4195(2018), ACSNano.13.7871(2019), PhysRevApplied.11.024061(2019)}, the increment of the supercurrent noise power with \(V_\mathrm{G}\) contradicts to the theoretical prediction that the integrated supercurrent noise should decrease as \(I_c^2\) \cite{PhysRevLett.76.3814(1996)}. Such an observation leads us to the conclusion that the gated nanobridge is overheated with respect to the environment, \textit{i.e.}, the observed suppression of the supercurrent can be explained by local overheating.
        
        In order to quantify the noise spectra at different gate voltages, we fitted them with a functional form \(Af^{-k}\), where \(k\) represents the slope of the spectra. The resulting dependence of the obtained exponent \(k\) as a function of \(V_\mathrm{G}\) is shown in the inset of \cref{fig:Noise}. At low gate voltages \(|V_\mathrm{G}|<\SI{25}{\volt}\) the value of \(k\) is close to zero, but it rapidly increases and approaches \(k\approx 1\) at higher voltages. If the gated nanobridge remains in thermal equilibrium, then, according to the fluctuation-dissipation theorem (FDT) \cite{LandauLifshitzVol5}, the observed growth of noise with the largely applied gate voltages can be explained either by a rise of the nanobridge temperature, or an increase in dissipation. Even if the system is driven out of thermodynamic equilibrium by the gate voltage, a generalised version of the FDT holds \cite{PhysRevLett.103.090601(2009)} and the conclusion about the higher temperature or greater dissipation still applies. Our noise measurements again confirm that the reported electrostatic field-effect in superconductors \cite{NatNano.13.802(2018), NanoLett.18.4195(2018), ACSNano.13.7871(2019), PhysRevApplied.11.024061(2019)} is caused by local overheating.
        
        Whether the system is in the equilibrium state or not can be qualitatively demonstrated via histograms of the measured noise \(\mathrm{I}\) and \(\mathrm{Q}\) components. \Cref{fig:histograms}(a) shows a normalised two-dimensional histogram at \(V_\mathrm{G}=\SI{50}{\volt}\) and its projections on the \(\mathrm{I}\) and \(\mathrm{Q}\) axes. The clear asymmetry of the histograms at high voltage is in strong contrast with the symmetric histograms at a low voltage \(V_\mathrm{G}=\SI{20}{\volt}\). The two projections of the histogram at \(V_\mathrm{G}=\SI{20}{\volt}\) are perfect Gaussian functions, which indicates that the system is in thermodynamic equilibrium \cite{LandauLifshitzVol5}. The distorted histograms at high gate voltage show that the equilibrium is broken. To determine the point at which the system is no longer in thermodynamic equilibrium, we quantify the non-Gaussianity using the 3\textsuperscript{rd} moment (skewness, \(\mu_3\)) of the distribution normalised to the standard deviation, \(\sigma_f\), \textit{viz.},  \(\mu_3/\sigma^3_f\). A non-zero value of the skewness shows that distribution is not  Gaussian, and system is out of thermal equilibrium. In addition to skewness, we present an analogue of the signal-to-noise ratio, \textit{viz.}, normalised frequency shift \(\Delta f/\sigma_f\). The gate dependencies of both quantities presented \cref{fig:histograms}(b). Both panels show that at low gate voltages \(|V_\mathrm{G}|<\SI{25}{\volt}\) the system stays in thermodynamic equilibrium. Outside that range the normalised frequency shift increases and skewness becomes negative. This is also supports our conclusions about local overheating of the nanobridge under the applied high gate voltage, which drives the system out of thermal equilibrium.
        
        \begin{figure}
            \includegraphics[width=\linewidth]{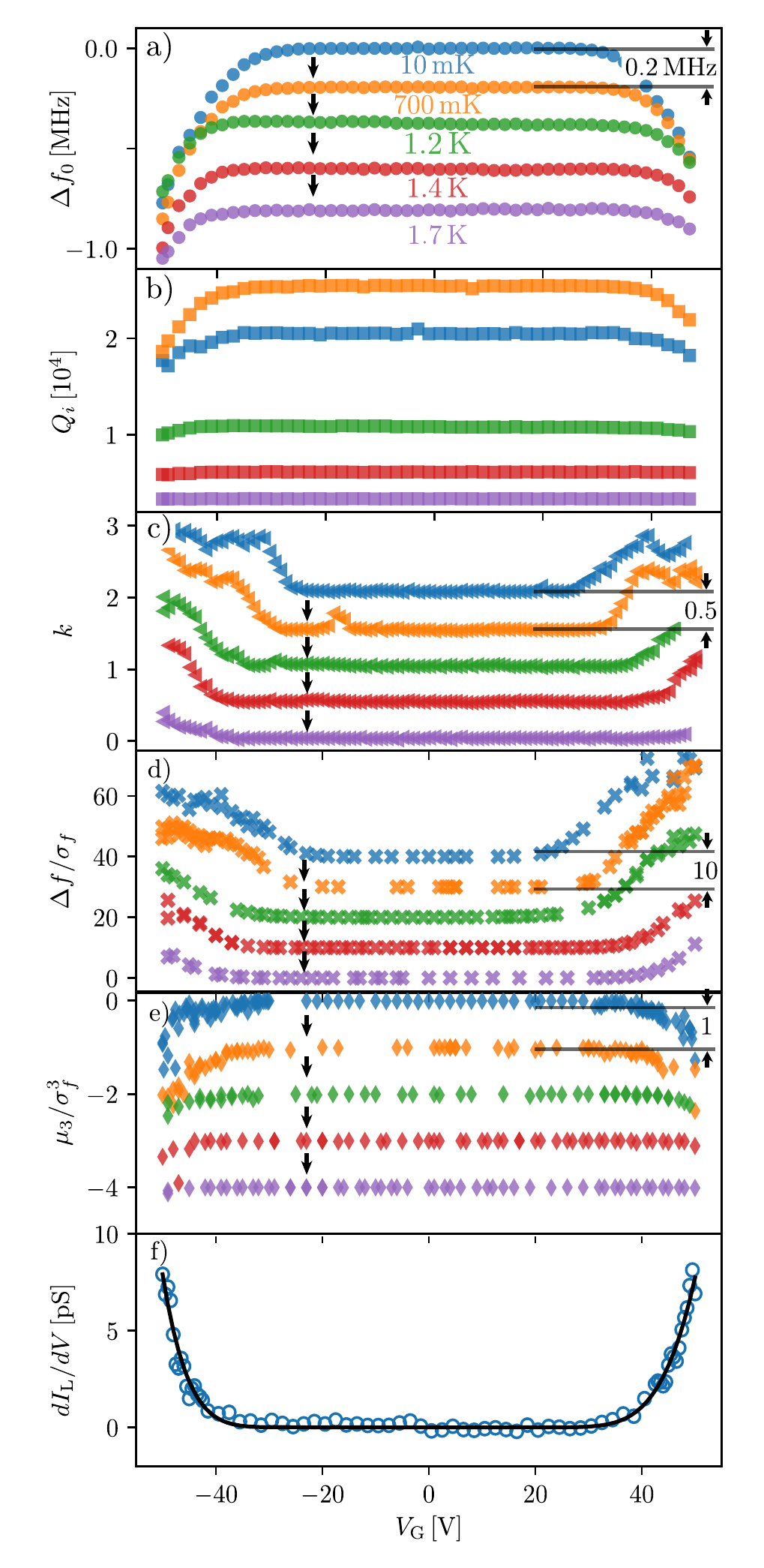}
            \caption{\label{fig:Summary} (Colour online) \textbf{Summary of the gate dependencies for all measured quantities at different temperatures.} Colours represent the cryostat temperature, which is common across all panels. For clarity, curves have been equidistantly shifted from each other, shown by the black arrows, by the value shown in black on the right of the corresponding panel. \textbf{(a)} The shift of the resonance frequency; \textbf{(b)} internal quality factor; \textbf{(c)} exponent \(k\) of the noise PSD \(\propto f^{-k}\); \textbf{(d)} normalised frequency shift and \textbf{(e)} normalised 3\textsuperscript{rd} moment (skewness) of the frequency fluctuations. \textbf{(f)} Measured differential conductance \(\dd I_\mathrm{L}/\dd V_\mathrm{G}\) of the leakage between the gate and Dayem bridge overlapped with a modelled curved obtained using \cref{eq:FN}.}
        \end{figure}
        
        The nonequilibrium distribution of quasiparticles should be less pronounced at higher temperatures due to the stronger electron-phonon coupling \cite{PhysRevB.49.5942(1994)} and our experiments at different temperatures confirm this. \Cref{fig:Summary} summarises the gate voltage dependence of all obtained quantities at different temperatures. All the six panels have a common feature, a plateau, in the range of about \(\pm\SI{25}{\volt}\) at \(\SI{10}{\milli\kelvin}\) with no dependence on \(V_\mathrm{G}\). This plateau becomes wider at higher temperatures. It is worth noting that the dependence of \(\Delta f_{0}\) and \(Q_i\) on \(V_\mathrm{G}\) shown in \cref{fig:Summary}(a) and (b), respectively, bare resemblance of the dependence of the critical current reported in earlier works (see, \textit{e.g.}, Fig.\,2(b) in Ref.\cite{NatNano.13.802(2018)} and Fig.\,2(b) in Ref. \cite{NanoLett.18.4195(2018)}). A similar behaviour, in which the plateau region widens with temperature is also observed for the exponent \(k\), the normalised frequency shift \(\Delta f/\sigma_f\) and the normalised skewness \(\mu_3/\sigma_f^3\), shown in \cref{fig:Summary}(c), (d) and (e), respectively. Such a dependence on temperature suggests that the observed effect is governed by heating of the constriction as higher gate voltage is required in order to see a change in the measured quantities at higher temperatures. All the data presented here point to the existence of an additional source of dissipation that results in higher noise and a distorted distribution function at higher \(V_\mathrm{G}\). The origin of such behaviour is attributed to leakage between gate and constriction. \Cref{fig:Summary}(f) presents the measured gate dependence of the leakage differential conductance, which has an onset at about \(\pm\SI{30}{\volt}\). The leakage of a similar magnitude was also reported in previous publications \cite{NatNano.13.802(2018), NanoLett.18.4195(2018), ACSNano.13.7871(2019), PhysRevApplied.11.024061(2019)}. This leakage depends on the gate voltage similarly to the dependence of the frequency shift, quality factor, noise, \textit{etc.}, \textit{i.e.}, there is a small effect below about \(|V_\mathrm{G}|\sim \SI{30}{\volt}\) followed by a stronger dependence above this value. This indicates that there may be a correlation between the gate leakage and all other observations.

        It is well known that electrons, when they are given enough energy, can escape from a metal surface in a process known as electron emission \cite{Modinos_FieldEmission}. One possible mechanism is escape under an intense electric field, \textit{viz.}, the field emission, where the supplied energy by the field is greater than the barrier height or workfunction, which have typical values of a few \SI{}{\electronvolt}. The gate voltages applied in our experiments and in \cite{NatNano.13.802(2018), NanoLett.18.4195(2018), ACSNano.13.7871(2019), PhysRevApplied.11.024061(2019)} reach the values of tens of volts in the distances of about \SI{100}{\nano\meter} and correspond to an electric field of \(\sim \SI{0.5}{\mega\volt\meter^{-1}}\), which is comparable with the breakdown electric field in vacuum gaps of \(\sim \SI{1}{\mega\volt\meter^{-1}}\) \cite{IEEETransDielElectIsul.20.1467(2013)}. The superconducting state of electrodes does not play any role as the characteristic superconducting energy gaps are of the order of \SI{1}{\milli\electronvolt} or below. 
        
        The emission current \(I_\mathrm{L}\) as a function of the applied voltage \(V_\mathrm{G}\) can be approximated within the Fowler-Nordheim model as \cite{ProcRoyalSocA.119.173(1928)}
        \begin{equation}
           I_\mathrm{L} = \alpha V_{\mathrm{G}}^2\exp\left(\dfrac{\beta}{V_\mathrm{G}}\right),
          \label{eq:FN}
        \end{equation}
        where \(\alpha\) and \(\beta\) are fitting parameters that depend on material properties and geometry. We use this expression to fit the experimentally found differential conductance, \(\dd I_\mathrm{L}/\dd V_\mathrm{G}\), of the gap between the gate and the nanobridge. \Cref{fig:Summary}(f) shows that the agreement between the experimental data and fit is excellent. We have also used \cref{eq:FN} to fit the the current-voltage characteristics presented in \cite{NatNano.13.802(2018), NanoLett.18.4195(2018)}. Again, an excellent agreement was found. This supports the idea that the electron field emission may be responsible for the observed effects. Indeed, each electron arriving at the nanoconstriction carries energy of tens of \SI{}{\electronvolt}, which is sufficient to destroy tens of thousands of Cooper pairs and generate quasiparticles, which is equivalent to thermal heating. From the leakage current \(\sim \SI{10}{\pico\ampere}\) at tens of volts, the average dissipated power at the nanoconstriction can be well above \SI{100}{\pico\watt}, which is sufficient to raise the effective temperature to \SI{1}{\kelvin} and above.
        
        Our experiment shows that there is a strong correlation between the gate voltage dependence of the leakage current and all other measured quantities such as the shift of the resonance frequency, the quality factor and the noise properties of the system. Therefore, based upon the presented evidences, we conclude that the suppression of the critical current in Dayem bridges reported in the earlier works \cite{NatNano.13.802(2018), NanoLett.18.4195(2018), ACSNano.13.7871(2019), PhysRevApplied.11.024061(2019)} also caused by local Joule heating generated by the leakage current at intense electric fields. While electron emission is claimed to be ruled out in the recent work \cite{arXiv:2006.07091}, the explanation is also supported by the recent experimental works \cite{arXiv:2005.00462, arXiv:2005.00584} and agrees with the existing theories of superconductivity.
        
        \textbf{Acknowledgements}
            We thank D.\,V.\,Averin, A.\,Braggio, F.\,Giazotto, I.\,Khaymovich, O.\,Kolosov, A.\,S.\,Melnikov, A.\,D.\,Zaikin and all members of the Lancaster University ULT group for fruitful discussions. This research was supported by the UK EPSRC Grants No. EP/P022197/1, EP/P024203/1, the Royal Society International Exchanges scheme (grant IES\textbackslash R3\textbackslash170054) and the EU H2020 European Microkelvin Platform (Grant Agreement 824109). This work was partially funded by the Joint Research Project PARAWAVE of the European Metrology Programme for Innovation and Research co-financed by the participating states and from the European Union’s Horizon 2020 research and innovation program.
        
        \textbf{Authors' contributions}
            The experiment was designed by SK and YuAP. The samples were fabricated by SK, AG and IG. Low-temperature measurements were carried out by AG, IG and SK. Data analysis was done by AG, IG, SK and VT. The interpretation of the results performed by SK, YuAP, AG, IG and VT. The manuscript was mainly written by AG, SK, YuAP and VT.
        
        \textbf{Online content}   
            Statements of data and code availability are available at \url{http://dx.doi.org/10.17635/lancaster/researchdata/xxx}.
        \bibliographystyle{apsrev4-1}
        \bibliography{bibliography}

\begin{thebibliography}{30}%
\makeatletter
\providecommand \@ifxundefined [1]{%
 \@ifx{#1\undefined}
}%
\providecommand \@ifnum [1]{%
 \ifnum #1\expandafter \@firstoftwo
 \else \expandafter \@secondoftwo
 \fi
}%
\providecommand \@ifx [1]{%
 \ifx #1\expandafter \@firstoftwo
 \else \expandafter \@secondoftwo
 \fi
}%
\providecommand \natexlab [1]{#1}%
\providecommand \enquote  [1]{``#1''}%
\providecommand \bibnamefont  [1]{#1}%
\providecommand \bibfnamefont [1]{#1}%
\providecommand \citenamefont [1]{#1}%
\providecommand \href@noop [0]{\@secondoftwo}%
\providecommand \href [0]{\begingroup \@sanitize@url \@href}%
\providecommand \@href[1]{\@@startlink{#1}\@@href}%
\providecommand \@@href[1]{\endgroup#1\@@endlink}%
\providecommand \@sanitize@url [0]{\catcode `\\12\catcode `\$12\catcode
  `\&12\catcode `\#12\catcode `\^12\catcode `\_12\catcode `\%12\relax}%
\providecommand \@@startlink[1]{}%
\providecommand \@@endlink[0]{}%
\providecommand \url  [0]{\begingroup\@sanitize@url \@url }%
\providecommand \@url [1]{\endgroup\@href {#1}{\urlprefix }}%
\providecommand \urlprefix  [0]{URL }%
\providecommand \Eprint [0]{\href }%
\providecommand \doibase [0]{https://doi.org/}%
\providecommand \selectlanguage [0]{\@gobble}%
\providecommand \bibinfo  [0]{\@secondoftwo}%
\providecommand \bibfield  [0]{\@secondoftwo}%
\providecommand \translation [1]{[#1]}%
\providecommand \BibitemOpen [0]{}%
\providecommand \bibitemStop [0]{}%
\providecommand \bibitemNoStop [0]{.\EOS\space}%
\providecommand \EOS [0]{\spacefactor3000\relax}%
\providecommand \BibitemShut  [1]{\csname bibitem#1\endcsname}%
\let\auto@bib@innerbib\@empty
\bibitem [{\citenamefont {Datta}(1995)}]{Datta_ElectronicTransport}%
  \BibitemOpen
  \bibfield  {author} {\bibinfo {author} {\bibfnamefont {S.}~\bibnamefont
  {Datta}},\ }\href {https://doi.org/10.1017/CBO9780511805776} {\emph {\bibinfo
  {title} {Electronic transport in mesoscopic systems}}}\ (\bibinfo
  {publisher} {Cambridge University Press},\ \bibinfo {year}
  {1995})\BibitemShut {NoStop}%
\bibitem [{\citenamefont {Sze}\ and\ \citenamefont
  {Ng}(2006)}]{SzeNg_PhysicsSemiconductorDevices}%
  \BibitemOpen
  \bibfield  {author} {\bibinfo {author} {\bibfnamefont {S.~M.}\ \bibnamefont
  {Sze}}\ and\ \bibinfo {author} {\bibfnamefont {K.~K.}\ \bibnamefont {Ng}},\
  }\href {https://doi.org/10.1002/0470068329} {\emph {\bibinfo {title} {Physics
  of Semiconductor Devices}}},\ \bibinfo {edition} {3rd}\ ed.\ (\bibinfo
  {publisher} {John Wiley \& Sons, Ltd},\ \bibinfo {year} {2006})\BibitemShut
  {NoStop}%
\bibitem [{\citenamefont {Ashcroft}\ and\ \citenamefont
  {Mermin}(1976)}]{Ashcroft_SolidStatePhysics}%
  \BibitemOpen
  \bibfield  {author} {\bibinfo {author} {\bibfnamefont {N.~W.}\ \bibnamefont
  {Ashcroft}}\ and\ \bibinfo {author} {\bibfnamefont {N.~D.}\ \bibnamefont
  {Mermin}},\ }\href@noop {} {\emph {\bibinfo {title} {Solid State Physics}}}\
  (\bibinfo  {publisher} {Cengage Learning, Inc},\ \bibinfo {year}
  {1976})\BibitemShut {NoStop}%
\bibitem [{\citenamefont {De~Simoni}\ \emph {et~al.}(2018)\citenamefont
  {De~Simoni}, \citenamefont {Paolucci}, \citenamefont {Solinas}, \citenamefont
  {Strambini},\ and\ \citenamefont {Giazotto}}]{NatNano.13.802(2018)}%
  \BibitemOpen
  \bibfield  {author} {\bibinfo {author} {\bibfnamefont {G.}~\bibnamefont
  {De~Simoni}}, \bibinfo {author} {\bibfnamefont {F.}~\bibnamefont {Paolucci}},
  \bibinfo {author} {\bibfnamefont {P.}~\bibnamefont {Solinas}}, \bibinfo
  {author} {\bibfnamefont {E.}~\bibnamefont {Strambini}},\ and\ \bibinfo
  {author} {\bibfnamefont {F.}~\bibnamefont {Giazotto}},\ }\href
  {https://doi.org/10.1038/s41565-018-0190-3} {\bibfield  {journal} {\bibinfo
  {journal} {Nature Nanotechnol.}\ }\textbf {\bibinfo {volume} {13}},\ \bibinfo
  {pages} {802} (\bibinfo {year} {2018})}\BibitemShut {NoStop}%
\bibitem [{\citenamefont {Paolucci}\ \emph {et~al.}(2018)\citenamefont
  {Paolucci}, \citenamefont {De~Simoni}, \citenamefont {Strambini},
  \citenamefont {Solinas},\ and\ \citenamefont
  {Giazotto}}]{NanoLett.18.4195(2018)}%
  \BibitemOpen
  \bibfield  {author} {\bibinfo {author} {\bibfnamefont {F.}~\bibnamefont
  {Paolucci}}, \bibinfo {author} {\bibfnamefont {G.}~\bibnamefont {De~Simoni}},
  \bibinfo {author} {\bibfnamefont {E.}~\bibnamefont {Strambini}}, \bibinfo
  {author} {\bibfnamefont {P.}~\bibnamefont {Solinas}},\ and\ \bibinfo {author}
  {\bibfnamefont {F.}~\bibnamefont {Giazotto}},\ }\href
  {https://doi.org/10.1021/acs.nanolett.8b01010} {\bibfield  {journal}
  {\bibinfo  {journal} {Nano Lett.}\ }\textbf {\bibinfo {volume} {18}},\
  \bibinfo {pages} {4195} (\bibinfo {year} {2018})}\BibitemShut {NoStop}%
\bibitem [{\citenamefont {De~Simoni}\ \emph {et~al.}(2019)\citenamefont
  {De~Simoni}, \citenamefont {Paolucci}, \citenamefont {Puglia},\ and\
  \citenamefont {Giazotto}}]{ACSNano.13.7871(2019)}%
  \BibitemOpen
  \bibfield  {author} {\bibinfo {author} {\bibfnamefont {G.}~\bibnamefont
  {De~Simoni}}, \bibinfo {author} {\bibfnamefont {F.}~\bibnamefont {Paolucci}},
  \bibinfo {author} {\bibfnamefont {C.}~\bibnamefont {Puglia}},\ and\ \bibinfo
  {author} {\bibfnamefont {F.}~\bibnamefont {Giazotto}},\ }\href
  {https://doi.org/10.1021/acsnano.9b02209} {\bibfield  {journal} {\bibinfo
  {journal} {ACS Nano}\ }\textbf {\bibinfo {volume} {13}},\ \bibinfo {pages}
  {7871} (\bibinfo {year} {2019})}\BibitemShut {NoStop}%
\bibitem [{\citenamefont {Paolucci}\ \emph {et~al.}(2019)\citenamefont
  {Paolucci}, \citenamefont {De~Simoni}, \citenamefont {Solinas}, \citenamefont
  {Strambini}, \citenamefont {Ligato}, \citenamefont {Virtanen}, \citenamefont
  {Braggio},\ and\ \citenamefont {Giazotto}}]{PhysRevApplied.11.024061(2019)}%
  \BibitemOpen
  \bibfield  {author} {\bibinfo {author} {\bibfnamefont {F.}~\bibnamefont
  {Paolucci}}, \bibinfo {author} {\bibfnamefont {G.}~\bibnamefont {De~Simoni}},
  \bibinfo {author} {\bibfnamefont {P.}~\bibnamefont {Solinas}}, \bibinfo
  {author} {\bibfnamefont {E.}~\bibnamefont {Strambini}}, \bibinfo {author}
  {\bibfnamefont {N.}~\bibnamefont {Ligato}}, \bibinfo {author} {\bibfnamefont
  {P.}~\bibnamefont {Virtanen}}, \bibinfo {author} {\bibfnamefont
  {A.}~\bibnamefont {Braggio}},\ and\ \bibinfo {author} {\bibfnamefont
  {F.}~\bibnamefont {Giazotto}},\ }\href
  {https://doi.org/10.1103/PhysRevApplied.11.024061} {\bibfield  {journal}
  {\bibinfo  {journal} {Phys. Rev. Appl.}\ }\textbf {\bibinfo {volume} {11}},\
  \bibinfo {pages} {024061} (\bibinfo {year} {2019})}\BibitemShut {NoStop}%
\bibitem [{\citenamefont {Fowler}\ and\ \citenamefont
  {Nordheim}(1928)}]{ProcRoyalSocA.119.173(1928)}%
  \BibitemOpen
  \bibfield  {author} {\bibinfo {author} {\bibfnamefont {R.~H.}\ \bibnamefont
  {Fowler}}\ and\ \bibinfo {author} {\bibfnamefont {L.}~\bibnamefont
  {Nordheim}},\ }\href {https://doi.org/10.1098/rspa.1928.0091} {\bibfield
  {journal} {\bibinfo  {journal} {P. R. Soc. Lond. A-Conta.}\ }\textbf
  {\bibinfo {volume} {119}},\ \bibinfo {pages} {173} (\bibinfo {year}
  {1928})}\BibitemShut {NoStop}%
\bibitem [{\citenamefont {Glover}\ and\ \citenamefont
  {Sherrill}(1960)}]{PhysRevLett.5.248(1960)}%
  \BibitemOpen
  \bibfield  {author} {\bibinfo {author} {\bibfnamefont {R.~E.}\ \bibnamefont
  {Glover}}\ and\ \bibinfo {author} {\bibfnamefont {M.~D.}\ \bibnamefont
  {Sherrill}},\ }\href {https://doi.org/10.1103/PhysRevLett.5.248} {\bibfield
  {journal} {\bibinfo  {journal} {Phys. Rev. Lett.}\ }\textbf {\bibinfo
  {volume} {5}},\ \bibinfo {pages} {248} (\bibinfo {year} {1960})}\BibitemShut
  {NoStop}%
\bibitem [{\citenamefont {Morpurgo}\ \emph {et~al.}(1998)\citenamefont
  {Morpurgo}, \citenamefont {Klapwijk},\ and\ \citenamefont {van
  Wees}}]{ApplPhysLett.72.966(1998)}%
  \BibitemOpen
  \bibfield  {author} {\bibinfo {author} {\bibfnamefont {A.~F.}\ \bibnamefont
  {Morpurgo}}, \bibinfo {author} {\bibfnamefont {T.~M.}\ \bibnamefont
  {Klapwijk}},\ and\ \bibinfo {author} {\bibfnamefont {B.~J.}\ \bibnamefont
  {van Wees}},\ }\href {https://doi.org/10.1063/1.120612} {\bibfield  {journal}
  {\bibinfo  {journal} {Appl. Phys. Lett.}\ }\textbf {\bibinfo {volume} {72}},\
  \bibinfo {pages} {966} (\bibinfo {year} {1998})}\BibitemShut {NoStop}%
\bibitem [{\citenamefont {Wilhelm}\ \emph {et~al.}(1998)\citenamefont
  {Wilhelm}, \citenamefont {Sch\"on},\ and\ \citenamefont
  {Zaikin}}]{PhysRevLett.81.1682(1998)}%
  \BibitemOpen
  \bibfield  {author} {\bibinfo {author} {\bibfnamefont {F.~K.}\ \bibnamefont
  {Wilhelm}}, \bibinfo {author} {\bibfnamefont {G.}~\bibnamefont {Sch\"on}},\
  and\ \bibinfo {author} {\bibfnamefont {A.~D.}\ \bibnamefont {Zaikin}},\
  }\href {https://doi.org/10.1103/PhysRevLett.81.1682} {\bibfield  {journal}
  {\bibinfo  {journal} {Phys. Rev. Lett.}\ }\textbf {\bibinfo {volume} {81}},\
  \bibinfo {pages} {1682} (\bibinfo {year} {1998})}\BibitemShut {NoStop}%
\bibitem [{\citenamefont {Clark}\ \emph {et~al.}(1980)\citenamefont {Clark},
  \citenamefont {Prance},\ and\ \citenamefont
  {Grassie}}]{JApplPhys.51.2736(1980)}%
  \BibitemOpen
  \bibfield  {author} {\bibinfo {author} {\bibfnamefont {T.~D.}\ \bibnamefont
  {Clark}}, \bibinfo {author} {\bibfnamefont {R.~J.}\ \bibnamefont {Prance}},\
  and\ \bibinfo {author} {\bibfnamefont {A.~D.~C.}\ \bibnamefont {Grassie}},\
  }\href {https://doi.org/10.1063/1.327935} {\bibfield  {journal} {\bibinfo
  {journal} {J. Appl. Phys.}\ }\textbf {\bibinfo {volume} {51}},\ \bibinfo
  {pages} {2736} (\bibinfo {year} {1980})}\BibitemShut {NoStop}%
\bibitem [{\citenamefont {Takayanagi}\ and\ \citenamefont
  {Kawakami}(1985)}]{PhysRevLett.54.2449(1985)}%
  \BibitemOpen
  \bibfield  {author} {\bibinfo {author} {\bibfnamefont {H.}~\bibnamefont
  {Takayanagi}}\ and\ \bibinfo {author} {\bibfnamefont {T.}~\bibnamefont
  {Kawakami}},\ }\href {https://doi.org/10.1103/PhysRevLett.54.2449} {\bibfield
   {journal} {\bibinfo  {journal} {Phys. Rev. Lett.}\ }\textbf {\bibinfo
  {volume} {54}},\ \bibinfo {pages} {2449} (\bibinfo {year}
  {1985})}\BibitemShut {NoStop}%
\bibitem [{\citenamefont {Nishino}\ \emph {et~al.}(1985)\citenamefont
  {Nishino}, \citenamefont {Miyake}, \citenamefont {Harada},\ and\
  \citenamefont {Kawabe}}]{IEEEElectDevLett.6.297(1985)}%
  \BibitemOpen
  \bibfield  {author} {\bibinfo {author} {\bibfnamefont {T.}~\bibnamefont
  {Nishino}}, \bibinfo {author} {\bibfnamefont {M.}~\bibnamefont {Miyake}},
  \bibinfo {author} {\bibfnamefont {Y.}~\bibnamefont {Harada}},\ and\ \bibinfo
  {author} {\bibfnamefont {U.}~\bibnamefont {Kawabe}},\ }\href
  {https://doi.org/10.1109/EDL.1985.26131} {\bibfield  {journal} {\bibinfo
  {journal} {IEEE Electr. Device L.}\ }\textbf {\bibinfo {volume} {6}},\
  \bibinfo {pages} {297} (\bibinfo {year} {1985})}\BibitemShut {NoStop}%
\bibitem [{\citenamefont {Kleinsasser}\ \emph {et~al.}(1989)\citenamefont
  {Kleinsasser}, \citenamefont {Jackson}, \citenamefont {McInturff},
  \citenamefont {Rammo}, \citenamefont {Pettit},\ and\ \citenamefont
  {Woodall}}]{ApplPhysLett.55.1909(1989)}%
  \BibitemOpen
  \bibfield  {author} {\bibinfo {author} {\bibfnamefont {A.~W.}\ \bibnamefont
  {Kleinsasser}}, \bibinfo {author} {\bibfnamefont {T.~N.}\ \bibnamefont
  {Jackson}}, \bibinfo {author} {\bibfnamefont {D.}~\bibnamefont {McInturff}},
  \bibinfo {author} {\bibfnamefont {F.}~\bibnamefont {Rammo}}, \bibinfo
  {author} {\bibfnamefont {G.~D.}\ \bibnamefont {Pettit}},\ and\ \bibinfo
  {author} {\bibfnamefont {J.~M.}\ \bibnamefont {Woodall}},\ }\href
  {https://doi.org/10.1063/1.102166} {\bibfield  {journal} {\bibinfo  {journal}
  {Appl. Phys. Lett.}\ }\textbf {\bibinfo {volume} {55}},\ \bibinfo {pages}
  {1909} (\bibinfo {year} {1989})}\BibitemShut {NoStop}%
\bibitem [{\citenamefont {Doh}\ \emph {et~al.}(2005)\citenamefont {Doh},
  \citenamefont {van Dam}, \citenamefont {Roest}, \citenamefont {Bakkers},
  \citenamefont {Kouwenhoven},\ and\ \citenamefont
  {De~Franceschi}}]{Science.309.272(2005)}%
  \BibitemOpen
  \bibfield  {author} {\bibinfo {author} {\bibfnamefont {Y.-J.}\ \bibnamefont
  {Doh}}, \bibinfo {author} {\bibfnamefont {J.~A.}\ \bibnamefont {van Dam}},
  \bibinfo {author} {\bibfnamefont {A.~L.}\ \bibnamefont {Roest}}, \bibinfo
  {author} {\bibfnamefont {E.~P. A.~M.}\ \bibnamefont {Bakkers}}, \bibinfo
  {author} {\bibfnamefont {L.~P.}\ \bibnamefont {Kouwenhoven}},\ and\ \bibinfo
  {author} {\bibfnamefont {S.}~\bibnamefont {De~Franceschi}},\ }\href
  {https://doi.org/10.1126/science.1113523} {\bibfield  {journal} {\bibinfo
  {journal} {Science}\ }\textbf {\bibinfo {volume} {309}},\ \bibinfo {pages}
  {272} (\bibinfo {year} {2005})}\BibitemShut {NoStop}%
\bibitem [{\citenamefont {Heersche}\ \emph {et~al.}(2007)\citenamefont
  {Heersche}, \citenamefont {Oostinga}, \citenamefont {Vandersypen},\ and\
  \citenamefont {Morpurgo}}]{Nature.446.56(2007)}%
  \BibitemOpen
  \bibfield  {author} {\bibinfo {author} {\bibfnamefont {P.}~\bibnamefont
  {Heersche}, \bibfnamefont {Hubert B.and Jarillo-Herrero}}, \bibinfo {author}
  {\bibfnamefont {J.~B.}\ \bibnamefont {Oostinga}}, \bibinfo {author}
  {\bibfnamefont {L.~M.~K.}\ \bibnamefont {Vandersypen}},\ and\ \bibinfo
  {author} {\bibfnamefont {A.~F.}\ \bibnamefont {Morpurgo}},\ }\href
  {https://doi.org/10.1038/nature05555} {\bibfield  {journal} {\bibinfo
  {journal} {Nature}\ }\textbf {\bibinfo {volume} {446}},\ \bibinfo {pages}
  {56} (\bibinfo {year} {2007})}\BibitemShut {NoStop}%
\bibitem [{\citenamefont {Takata}\ \emph {et~al.}(2015)\citenamefont {Takata},
  \citenamefont {Inagaki}, \citenamefont {Kawae}, \citenamefont {Ienaga},\ and\
  \citenamefont {Tsujii}}]{J.PhysicsConf.592.012137}%
  \BibitemOpen
  \bibfield  {author} {\bibinfo {author} {\bibfnamefont {H.}~\bibnamefont
  {Takata}}, \bibinfo {author} {\bibfnamefont {Y.}~\bibnamefont {Inagaki}},
  \bibinfo {author} {\bibfnamefont {T.}~\bibnamefont {Kawae}}, \bibinfo
  {author} {\bibfnamefont {K.}~\bibnamefont {Ienaga}},\ and\ \bibinfo {author}
  {\bibfnamefont {H.}~\bibnamefont {Tsujii}},\ }\href
  {https://doi.org/10.1088/1742-6596/592/1/012137} {\bibfield  {journal}
  {\bibinfo  {journal} {J. Phys. Conf. Ser.}\ }\textbf {\bibinfo {volume}
  {592}},\ \bibinfo {pages} {012137} (\bibinfo {year} {2015})}\BibitemShut
  {NoStop}%
\bibitem [{\citenamefont {Mohsin}(2017)}]{Mohsin_ManifoldMultiplexer}%
  \BibitemOpen
  \bibfield  {author} {\bibinfo {author} {\bibfnamefont {I.}~\bibnamefont
  {Mohsin}},\ }in\ \href {https://doi.org/10.5772/66407} {\emph {\bibinfo
  {booktitle} {Microwave Systems and Applications}}},\ \bibinfo {editor}
  {edited by\ \bibinfo {editor} {\bibfnamefont {S.}~\bibnamefont {Goudos}}}\
  (\bibinfo  {publisher} {IntechOpen},\ \bibinfo {year} {2017})\ Chap.~\bibinfo
  {chapter} {4}\BibitemShut {NoStop}%
\bibitem [{\citenamefont {Di~Paolo}(2000)}]{DiPaolo_NetworksDevices}%
  \BibitemOpen
  \bibfield  {author} {\bibinfo {author} {\bibfnamefont {F.}~\bibnamefont
  {Di~Paolo}},\ }\href {https://doi.org/10.1201/9781315220369} {\emph {\bibinfo
  {title} {Networks and Devices Using Planar Transmissions Lines}}}\ (\bibinfo
  {publisher} {CRC Press},\ \bibinfo {year} {2000})\BibitemShut {NoStop}%
\bibitem [{\citenamefont {Averin}\ and\ \citenamefont
  {Imam}(1996)}]{PhysRevLett.76.3814(1996)}%
  \BibitemOpen
  \bibfield  {author} {\bibinfo {author} {\bibfnamefont {D.}~\bibnamefont
  {Averin}}\ and\ \bibinfo {author} {\bibfnamefont {H.~T.}\ \bibnamefont
  {Imam}},\ }\href {https://doi.org/10.1103/PhysRevLett.76.3814} {\bibfield
  {journal} {\bibinfo  {journal} {Phys. Rev. Lett.}\ }\textbf {\bibinfo
  {volume} {76}},\ \bibinfo {pages} {3814} (\bibinfo {year}
  {1996})}\BibitemShut {NoStop}%
\bibitem [{\citenamefont {Mart\'{\i}n-Rodero}\ \emph
  {et~al.}(1996)\citenamefont {Mart\'{\i}n-Rodero}, \citenamefont {Yeyati},\
  and\ \citenamefont {Garc\'{\i}a-Vidal}}]{PhysRevB.53.R8891(1996)}%
  \BibitemOpen
  \bibfield  {author} {\bibinfo {author} {\bibfnamefont {A.}~\bibnamefont
  {Mart\'{\i}n-Rodero}}, \bibinfo {author} {\bibfnamefont {A.~L.}\ \bibnamefont
  {Yeyati}},\ and\ \bibinfo {author} {\bibfnamefont {F.~J.}\ \bibnamefont
  {Garc\'{\i}a-Vidal}},\ }\href {https://doi.org/10.1103/PhysRevB.53.R8891}
  {\bibfield  {journal} {\bibinfo  {journal} {Phys. Rev. B}\ }\textbf {\bibinfo
  {volume} {53}},\ \bibinfo {pages} {R8891} (\bibinfo {year}
  {1996})}\BibitemShut {NoStop}%
\bibitem [{\citenamefont {Landau}\ and\ \citenamefont
  {Lifshitz}(1980)}]{LandauLifshitzVol5}%
  \BibitemOpen
  \bibfield  {author} {\bibinfo {author} {\bibfnamefont {L.~D.}\ \bibnamefont
  {Landau}}\ and\ \bibinfo {author} {\bibfnamefont {E.~M.}\ \bibnamefont
  {Lifshitz}},\ }\href
  {https://www.elsevier.com/books/statistical-physics/landau/978-0-08-057046-4}
  {\emph {\bibinfo {title} {Statistical Physics, Part 1}}},\ \bibinfo {edition}
  {3rd}\ ed.,\ Course of Theoretical Physics, Volume 5\ (\bibinfo  {publisher}
  {Butterworth-Heinemann},\ \bibinfo {year} {1980})\BibitemShut {NoStop}%
\bibitem [{\citenamefont {Prost}\ \emph {et~al.}(2009)\citenamefont {Prost},
  \citenamefont {Joanny},\ and\ \citenamefont
  {Parrondo}}]{PhysRevLett.103.090601(2009)}%
  \BibitemOpen
  \bibfield  {author} {\bibinfo {author} {\bibfnamefont {J.}~\bibnamefont
  {Prost}}, \bibinfo {author} {\bibfnamefont {J.-F.}\ \bibnamefont {Joanny}},\
  and\ \bibinfo {author} {\bibfnamefont {J.~M.~R.}\ \bibnamefont {Parrondo}},\
  }\href {https://doi.org/10.1103/PhysRevLett.103.090601} {\bibfield  {journal}
  {\bibinfo  {journal} {Phys. Rev. Lett.}\ }\textbf {\bibinfo {volume} {103}},\
  \bibinfo {pages} {090601} (\bibinfo {year} {2009})}\BibitemShut {NoStop}%
\bibitem [{\citenamefont {Wellstood}\ \emph {et~al.}(1994)\citenamefont
  {Wellstood}, \citenamefont {Urbina},\ and\ \citenamefont
  {Clarke}}]{PhysRevB.49.5942(1994)}%
  \BibitemOpen
  \bibfield  {author} {\bibinfo {author} {\bibfnamefont {F.~C.}\ \bibnamefont
  {Wellstood}}, \bibinfo {author} {\bibfnamefont {C.}~\bibnamefont {Urbina}},\
  and\ \bibinfo {author} {\bibfnamefont {J.}~\bibnamefont {Clarke}},\ }\href
  {https://doi.org/10.1103/PhysRevB.49.5942} {\bibfield  {journal} {\bibinfo
  {journal} {Phys. Rev. B}\ }\textbf {\bibinfo {volume} {49}},\ \bibinfo
  {pages} {5942} (\bibinfo {year} {1994})}\BibitemShut {NoStop}%
\bibitem [{\citenamefont {Modinos}(1984)}]{Modinos_FieldEmission}%
  \BibitemOpen
  \bibfield  {author} {\bibinfo {author} {\bibfnamefont {A.}~\bibnamefont
  {Modinos}},\ }\href {https://doi.org/10.1007/978-1-4757-1448-7} {\emph
  {\bibinfo {title} {Field, Thermionic, and Secondary Electron Emission
  Spectroscopy}}}\ (\bibinfo  {publisher} {Springer US},\ \bibinfo {year}
  {1984})\BibitemShut {NoStop}%
\bibitem [{\citenamefont {Lyon}\ and\ \citenamefont
  {Hubler}(2013)}]{IEEETransDielElectIsul.20.1467(2013)}%
  \BibitemOpen
  \bibfield  {author} {\bibinfo {author} {\bibfnamefont {D.}~\bibnamefont
  {Lyon}}\ and\ \bibinfo {author} {\bibfnamefont {A.}~\bibnamefont {Hubler}},\
  }\href {https://doi.org/10.1109/TDEI.2013.6571470} {\bibfield  {journal}
  {\bibinfo  {journal} {IEEE T. Dielect El. In.}\ }\textbf {\bibinfo {volume}
  {20}},\ \bibinfo {pages} {1467} (\bibinfo {year} {2013})}\BibitemShut
  {NoStop}%
\bibitem [{\citenamefont {Rocci}\ \emph {et~al.}(2020)\citenamefont {Rocci},
  \citenamefont {Simoni}, \citenamefont {Puglia}, \citenamefont {Esposti},
  \citenamefont {Strambini}, \citenamefont {Zannier}, \citenamefont {Sorba},\
  and\ \citenamefont {Giazotto}}]{arXiv:2006.07091}%
  \BibitemOpen
  \bibfield  {author} {\bibinfo {author} {\bibfnamefont {M.}~\bibnamefont
  {Rocci}}, \bibinfo {author} {\bibfnamefont {G.~D.}\ \bibnamefont {Simoni}},
  \bibinfo {author} {\bibfnamefont {C.}~\bibnamefont {Puglia}}, \bibinfo
  {author} {\bibfnamefont {D.~D.}\ \bibnamefont {Esposti}}, \bibinfo {author}
  {\bibfnamefont {E.}~\bibnamefont {Strambini}}, \bibinfo {author}
  {\bibfnamefont {V.}~\bibnamefont {Zannier}}, \bibinfo {author} {\bibfnamefont
  {L.}~\bibnamefont {Sorba}},\ and\ \bibinfo {author} {\bibfnamefont
  {F.}~\bibnamefont {Giazotto}},\ }\href@noop {} {\bibinfo {title}
  {Gate-controlled suspended titanium nanobridge supercurrent transistor}}
  (\bibinfo {year} {2020}),\ \Eprint {https://arxiv.org/abs/2006.07091}
  {arXiv:2006.07091} \BibitemShut {NoStop}%
\bibitem [{\citenamefont {Ritter}\ \emph {et~al.}(2020)\citenamefont {Ritter},
  \citenamefont {Fuhrer}, \citenamefont {Haxell}, \citenamefont {Hart},
  \citenamefont {Gumann}, \citenamefont {Riel},\ and\ \citenamefont
  {Nichele}}]{arXiv:2005.00462}%
  \BibitemOpen
  \bibfield  {author} {\bibinfo {author} {\bibfnamefont {M.~F.}\ \bibnamefont
  {Ritter}}, \bibinfo {author} {\bibfnamefont {A.}~\bibnamefont {Fuhrer}},
  \bibinfo {author} {\bibfnamefont {D.~Z.}\ \bibnamefont {Haxell}}, \bibinfo
  {author} {\bibfnamefont {S.}~\bibnamefont {Hart}}, \bibinfo {author}
  {\bibfnamefont {P.}~\bibnamefont {Gumann}}, \bibinfo {author} {\bibfnamefont
  {H.}~\bibnamefont {Riel}},\ and\ \bibinfo {author} {\bibfnamefont
  {F.}~\bibnamefont {Nichele}},\ }\href@noop {} {\bibinfo {title} {A
  superconducting switch actuated by injection of high energy electrons}}
  (\bibinfo {year} {2020}),\ \Eprint {https://arxiv.org/abs/2005.00462}
  {arXiv:2005.00462} \BibitemShut {NoStop}%
\bibitem [{\citenamefont {Alegria}\ \emph {et~al.}(2020)\citenamefont
  {Alegria}, \citenamefont {Bøttcher}, \citenamefont {Saydjari}, \citenamefont
  {Pierce}, \citenamefont {Lee}, \citenamefont {Harvey}, \citenamefont {Vool},\
  and\ \citenamefont {Yacoby}}]{arXiv:2005.00584}%
  \BibitemOpen
  \bibfield  {author} {\bibinfo {author} {\bibfnamefont {L.~D.}\ \bibnamefont
  {Alegria}}, \bibinfo {author} {\bibfnamefont {C.~G.}\ \bibnamefont
  {Bøttcher}}, \bibinfo {author} {\bibfnamefont {A.~K.}\ \bibnamefont
  {Saydjari}}, \bibinfo {author} {\bibfnamefont {A.~T.}\ \bibnamefont
  {Pierce}}, \bibinfo {author} {\bibfnamefont {S.~H.}\ \bibnamefont {Lee}},
  \bibinfo {author} {\bibfnamefont {S.~P.}\ \bibnamefont {Harvey}}, \bibinfo
  {author} {\bibfnamefont {U.}~\bibnamefont {Vool}},\ and\ \bibinfo {author}
  {\bibfnamefont {A.}~\bibnamefont {Yacoby}},\ }\href@noop {} {\bibinfo {title}
  {High-energy quasiparticle injection in mesoscopic superconductors}}
  (\bibinfo {year} {2020}),\ \Eprint {https://arxiv.org/abs/2005.00584}
  {arXiv:2005.00584} \BibitemShut {NoStop}%
\end{thebibliography}%
\end{document}